\documentclass[12pt]{article}

\catcode`\@=11
\@addtoreset{equation}{section}

\global\arraycolsep=2pt
\usepackage{amsbsy,amssymb,latexsym,amsfonts,amsmath}
\usepackage{graphicx,color}

\allowdisplaybreaks

\begin{document}
\begin{flushright}
\parbox{4.2cm}
{UCB-PTH-10/07}
\end{flushright}

\vspace*{0.7cm}

\begin{center}
{\Large \bf 
Gravity Dual for a Model of Perception}
\vspace*{2.0cm}\\
{Yu Nakayama}
\end{center}
\vspace*{-0.2cm}
\begin{center}
{\it Berkeley Center for Theoretical Physics, \\ 
University of California, Berkeley, CA 94720, USA }
\vspace{4.8cm}
\end{center}

\begin{abstract} 
One of the salient features of human perception is its invariance under dilatation in addition to the Euclidean group, but its non-invariance under special conformal transformation. We investigate a holographic approach to the information processing in image discrimination with this feature. 
We claim that a strongly coupled analogue of the statistical model proposed by Bialek and Zee can be holographically realized in scale invariant but non-conformal Euclidean geometries. We identify the Bayesian probability distribution of our generalized Bialek-Zee model with the 
GKPW partition function of the dual gravitational system.
We provide a concrete example of the geometric configuration based on a vector condensation model coupled with the Euclidean Einstein-Hilbert action. From the proposed geometry, we study sample correlation functions to compute the Bayesian probability distribution.

\end{abstract}

\thispagestyle{empty} 

\setcounter{page}{0}

\newpage

\section{Introduction} 
One of the salient features of human perception is its invariance under various symmetries. A triangle is recognized as a triangle no matter how it is rotated, translated, or enlarged.\footnote{See e.g. \cite{Wiskott} and reference therein for the experimental evidence.} More importantly, however, it is {\it not} invariant under special conformal transformation: the conformally transformed images are recognized as distorted ones. It is this feeling of ``distorted" that makes our perception non-invariant under the special conformal transformation.

The symmetry principle has played a significant role in physics, and so must be in theories of perception. From the symmetry viewpoint, in \cite{Bialek:1986it}\cite{Bialek:1987qc}, Bialek and Zee presented a field theoretic model for invariant perception. Their field theory model is constructed to be invariant under the dilatation in addition to the Euclidean group, but not invariant under the special conformal transformation. The underlying reasoning is 
simple: the Euclidean group together with the dilatation is a key symmetry of the human perception, and the model must admit this symmetry in a manifest way. Their model is based on a free Gaussian field theory, but the use of the free field theory is no way mandatory from the symmetry. The aim of this paper is to depart from the free Gaussian field theory and to generalize their proposal in the strongly coupled limit by using the idea of holography. In other words, we would like to propose a holographic dual approach to the field theories for invariant perception.

The holographic approaches to the strongly coupled field theories have been successful in various contexts. We can name many examples from the QCD and condensed matter systems even to the quantum finance \cite{Nakayama:2009ww}. Given the unreasonable effectiveness of the holographic approaches so far in various realms of physics and in broader sciences, it is reasonable to assume that our complicated human perception is not an exception.
In our approach, we will show that the GKPW partition function \cite{Gubser:1998bc}\cite{Witten:1998qj} of the gravitational system is identified with the Bayesian probability distribution, which we will utilize to decode the original image from the percerived data that has been distorted by the exterior noise. Our recognition system computes the gravitational partition function at every moment!

The holographic approach, however, has one puzzling issue related to the existence of scale invariant but non-conformal field theories \cite{Coleman:1970je}\cite{Polchinski:1987dy}\cite{Dorigoni:2009ra}. We know a very few examples of such field theories \cite{Hull:1985rc}\cite{Riva:2005gd}\cite{Ho:2008nr}\cite{Nak}. It is even shown \cite{Zamolodchikov:1986gt}\cite{Polchinski:1987dy} that under some mild assumptions such as unitarity, Poincar\'e invariance and discreteness of the spectrum, it is impossible to construct one in (1+1)-dimension. Recall that the human perception deals with the two-dimensional screen, so the associated field theory used in the Bialek-Zee model is two-dimensional, and the situation is very close (upon the Wick rotation). Furthermore, in \cite{Nakayama:2009qu}\cite{Nakayama:2009fe}, we showed that the gravity dual for the scale invariant but non-conformal field theory in the warped compactification cannot be realized whenever the null energy condition is satisfied.

We circumvent these obstacles both in field theories and in holographic gravitational systems by discarding the unitarity condition and the energy condition. 
First of all, the free field theory model proposed by Bialek and Zee does not satisfy the reflection positivity, which means that its Lorentzian continuation is non-unitary. We emphasize that the unitarity or reflection positivity is not essential in our setup that is intrinsically Euclidean unlike in the conventional Lorentzian quantum field theories. Analogously, we restrict ourselves to the Euclidean gravity and do not require the consistency of its Lorentzian continuation. As a consequence, there is no notion of energy condition in our Euclidean gravity action.

The organization of the paper is as follows. In section 2, we review the Bialek-Zee model of invariant perception. A simple application of the Bayesian statistics gives us a natural model of invariant perception written in the language of Euclidean field theory. Their model is based on the Gaussian free field theory, and our goal is to replace the free field theory with strongly coupled ones by using the holographic construction. In section 3, we propose a gravity dual approach to the invariant perception. In section 3.1, we study the geometry which is invariant under the scale transformation but not invariant under the special conformal transformation. In section 3.2, we show that the GKPW partition function of our Euclidean geometry is nothing but the Bayesian probability distribution for the invariant perception. In section 4, we give some sample holographic computations of the Bayesian probability distribution from the geometry we propose in this paper. In section 5, we present the discussions and summary.
We have two appendices: in appendix A, we summarize the distinction between the scale invariant field theories and the conformal invariant field theories by focusing on the behavior of the energy momentum tensor. In appendix B, we summarize relevant formulae for  the modified Bessel function used in section 4.

\section{Bialek-Zee model}

\subsection{Bayesian statistics and perception}
The Bayesian statistics plays a central role in constructing models of perception and recognition, and we would like to begin with a brief review of the subject. The formulation of the Bayesian statistics is in close parallel with the statistical mechanics in equilibrium and the Euclidean field theory in the continuum limit. The analogy will be useful when we discuss the holographic dual in later sections.

We first denote the probability to observe the data $y$ with a given set of parameters $x$ as
\begin{align}
L(y|x) = \frac{\exp\left(-S[y,x]\right)}{Z_L(x)} \ , 
\end{align}
where the partition function is given by $Z_L(x) = \sum_y \exp(-S[y,x])$. As a function of $x$, $L(y|x)$ is called ``likelihood" in Bayesian statistics.  In relation to the statistical mechanics, we regard the parameters $x$ as coupling constants or the source in the ``action" $S[y,x]$. The data $y$ are, on the other hand, regarded as a dynamical variable. In our application, the probabilistic variable $y$ becomes a Euclidean field, so the sum over $y$ is replaced by a functional integral or Euclidean path integral.

Bayesian statistics assumes a concept of the prior distribution for $x$ in order to estimate the bare parameters $x$ in terms of $y$. The prior distribution can be written as
\begin{align}
\pi(x) = \frac{\exp(-S_\pi(x))}{Z_\pi}  \ , \label{prior}
\end{align}
where the prior partition function is defined as $Z_\pi = \sum_x \exp(-S_\pi(x))$.
With the usage of the prior distribution, the Bayes theorem yields a probability of $x$, given the observed data $y$, as
\begin{align}
P(x|y) = \frac{L(y|x) \pi (x)}{\sum_x L(y|x) \pi (x)} \ . \label{post}
\end{align}
We can estimate the value of $x$ by using the probability $P(x|y)$ as we like (e.g. by taking the expectation value, by computing the median and so on). In particular, the strategy to estimate $x$ as the value of $x$ that maximize \eqref{post} is known as maximal a posteriori estimate.
The probability \eqref{post} is known as a posterior distribution in Bayesian statistics. 

In this paper, we will not discuss the prior distribution \eqref{prior} very much, but rather we mainly focus on the likelihood, which not only plays a central role in the Bayes theorem, but also plays an important role in understanding the complexity of the human perception. This is because the computational complexity typically lies in evaluating the likelihood $L(x|y)$ in the Bayes formula \eqref{post} rather than the prior distribution $\pi(x)$.
We note that the estimate from the Bayesian statistics cannot be free from a certain amount of subjectivity: we cannot exclude the ``unknown" prior distribution, and furthermore, we have to decide which criterion we employ to estimate the ``most plausible" value of $x$ from the probability $P(x|y)$.\footnote{Note that physics cannot always be free from the prior distribution (see e.g. \cite{Bayesian} for the applications in experimental high energy physics.). For instance, with no reference to the prior, we can directly compute the probability distribution of the scattering data from the parameters of the Lagrangian, say Higgs mass. We, however, require a prior distribution (i.e. the probability that the Higgs mass takes a particular value in the enemble of the universe) to estimate the Higgs mass from the experimental data within the framework of the Bayesian statistics.}

A model of perception or recognition of the image under the stochastic noise was proposed in \cite{Bialek:1986it}, and we would like to use their main concept to construct the holographic approach to the invariant perception.  Let us consider an image described by a scalar field $\phi(x)$, where $x$ denotes the two dimensional screen with our vision in mind.  For simplicity, we do not introduce the internal structure ({\it e.g.} color), and treat it as a monochrome picture. The Euclidean field  $\phi(x)$ therefore denotes the grey scale that we perceive at a point $x_i$ in the two-dimensional screen.
We believe that the image $\phi(x)$ is obtained from the ``original" image $\phi_0(x)$ by distorting this image and by adding noise.\footnote{This belief that the original exists can never be proven in human perception. Nevertheless, we continue to hold this philosophical assumption. The assumption is not only essential but practically always useful in any actual model of detection or image processing.} 

The central problem of perception or image processing is to guess $\phi_0(x)$ from the observed $\phi(x)$. 
Since we do not know how the noise is added and how the image is distorted, we are forced to think about the probability distribution (or more precisely liklihood in the Bayesian terminlogy) $L[\phi(x)|\phi_0(x)]$ that defines the conditional probability to observe $\phi(x)$ assuming that $\phi_0(x)$ is given. When the noise is parametrized by a classical system, it may be encoded in a random field variable $\chi(x)$. The probability distribution of the noise is described by its own distribution $P[\chi(x)] = \exp(-W[\chi(x)])$.
 The physical law determines the distribution functional of $\phi(x)$ with an initial data $\phi_0(x)$ and the noise $\chi(x)$ as $P[\phi(x)|\phi_0(x); \chi(x)]$. The accessible probability distribution is related to the underlying probability distribution by 
\begin{align}
L[\phi(x)|\phi_0(x)] = \frac{1}{Z[\phi_0(x)]} \int \mathcal{D} \chi(x) e^{-W[\chi(x)]} P[\phi(x)|\phi_0(x); \chi(x)] \ .
\end{align}
Here, the partition function $Z[\phi_0(x)]$ is defined as
\begin{align}
Z[\phi_0(x)] = e^{-F[\phi_0(x)]} = \int \mathcal{D}\phi(x) \exp\left(-S_{\text{eff}}[\phi(x);\phi_0(x)] \right) \ .
\end{align}
Therefore, the computation of the efficiency of the perception boils down to evaluating the effective action
\begin{align}
\exp\left(-S_{\text{eff}}[\phi(x);\phi_0(x)] \right)  = \int \mathcal{D} \chi(x) e^{-W[\chi(x)]} P[\phi(x)|\phi_0(x); \chi(x)] \ .
\end{align}

If one neglects the prior distribution, or if we assume a flat prior distribution with respect to the path integral measure $\mathcal{D}\phi_0(x)$, one can perform the optimal discrimination by using the maximal a posteriori estimate: the functional variation or ``quantum equation of motion"
\begin{align}
\frac{\delta (S_{\text{eff}}[\phi(x); \phi_0(x)]-F[\phi_0(x)])}{\delta \phi_0(x)} = 0 
\end{align}
gives us the most plausible estimate of the prior image $\phi_0(x)$ with an observed image $\phi(x)$ (under the flat prior distribution).

\subsection{A free field model}
In order to compute the effective action $S_{\text{eff}}[\phi(x);\phi_0(x)]$, we have to specify the probability distribution $P[\phi(x)|\phi_0(x); \chi(x)]
$ as well as the noise distribution $e^{-W[\chi(x)]}$. We will follow the model proposed in \cite{Bialek:1986it}. The first assumption in their model is that the noise is white, so we can write the effective action as
\begin{align}
\exp\left( -S_{\text{eff}} [\phi(x);\phi_0(x)] \right) = \int \mathcal{D} S(x) \exp\left( - W[S(x)] - \frac{1}{2C} \int d^2 x \left(\phi(x) -\phi_0(y(x))\right)^2 \right) \ ,
\end{align}
where the distortion is denoted as $y_i = x_i + S_i(x)$. The intensity of the noise is governed by $C^{-1}$.

The salient features of the human perception is that it is (approximately) invariant under translation and the Euclidean rotation as well as dilatation, but not invariant under special conformal transformation. The assumed invariance demands that the weighting functional $W[S(x)]$ for the distortion reflects the invariance under the Euclidean group and the dilatation. If one regards the integration over the functional space $\mathcal{D} S(x)$ with the weighting functional $W[S(x)]$ as defining a Euclidean field theory, the problem is equivalent to finding a scale invariant but non-conformal field theory on the Euclidean plane.

The simplest choice for $W[S(x)]$ is to use a free Gaussian field. The most general free action for the scale invariant but not conformal invariant vector field with the canonical dimension is given by
\begin{align}
W[S(x)] = \int d^2x \left( \frac{1}{4g_1^2} (\partial_i S_j - \partial_j S_i)^2+ \frac{1}{2g_2^2}(\partial_i S_i)^2 \right) \ . \label{rc}
\end{align}
We note that in the original model of \cite{Bialek:1986it}, the field $S_i$ is decomposed as $S_i = \partial_i \chi + \epsilon_{ij} \partial_j \phi$, which is always possible in two-dimensional field theories. 
The model is obviously invariant under the Euclidean group and dilatation, but it is not invariant under the special conformal transformation unless $g_1^2 = g_2^2$ \cite{Riva:2005gd} or $g_1^2= -g_2^2$ \cite{ElShowk:2011gz}. It is easy to see that the trace of the energy momentum tensor is a total divergence: $T^{i}_{\ i} =  \partial^iJ_i$, but the virial current $J_i$ is not a total derivative, so it is impossible to improve the energy momentum tensor so that it becomes traceless. See appendix A for more discussions.

 It is interesting to note that exactly the same model was studied in \cite{Riva:2005gd}, in the context of presenting a physical example of a scale invariant but non-conformal field theory. They studied the theory of elasticity in two-dimension. In their context, the tensor $\frac{1}{2}\left(\partial_i S_j + \partial_j S_i\right)$ is known as the strain tensor, and it is built out of the displacement field $S_i$. 

The two-point function for the action \eqref{rc} can be easily computed as
\begin{align}
\langle S_i(k) S_j(p) \rangle = \delta^{(2)}(k+p) \left( \frac{g_1^2}{k^2} \left(\delta_{ij}-\frac{k_ik_j}{k^2}\right) + \frac{g_2^2}{k^2} \frac{k_ik_j}{k^2}\right) \ .
\end{align}
This expression is consistent with the most general form of the two-point functions of the vector fields (with a scaling dimension $\Delta$) in scale invariant but not necessarily conformally invariant field theories in $d$-dimension:
\begin{align}
\langle O_i(k) O_j(p) \rangle = \delta^{(d)}(k+p) \left( C_1 \frac{\delta_{ij}}{k^{-2\Delta+d}} + C_2 \frac{k_ik_j}{k^{2-2\Delta+d}} \right) \ .
\end{align}
The two-point function is consistent with the conformal invariance if and only if $C_2/C_1 = -\frac{2\Delta-d}{\Delta-1}$. 
Accordingly, one cannot distinguish the free field model with the more complicated strongly coupled model at the level of two-point function, but they will differ in higher point functions. In particular, the higher cumulants are all zero for the free field theory, but they are not for the interacting theories.

We note that the Lorentzian continuation of the free field model \eqref{rc} is non-unitary except for $\frac{1}{g_2^2} = 0$ when we can impose the gauge symmetry. This is the reason why this model evades the Zamolodchikov-Polchinski theorem  \cite{Zamolodchikov:1986gt}\cite{Polchinski:1987dy} that states the equivalence between the scale invariance and conformal invariance. 

In the following section, we try to replace the free field action $W[S(x)]$ by abstract scale invariant field theories. In particular, we focus on the strongly coupled example where the path integral over $W[S(x)]$ is facilitated by the holographic dual computation. Since we do not know any good interacting examples of vector field theories that are scale invariant but not conformally invariant, our gravity dual approach is the first non-trivial example of such a non-Gaussian construction.

\section{Gravity dual for perception}

\subsection{Geometry and field configuration}
As we have discussed in section 2, the weighting functional $W[S(x)]$ in the Bayesian probability distributions may be arbitrarily complicated as long as it preserves the Euclidean group and the scale invariance, but not conformal invariance. We are, thus, able to replace the path integral for the weighting functional by an abstract scale invariant (but non-conformal) field theory. In that case, given its complexity, we may compute the Bayesian probability distribution by using the holographic technique.

First, we have to specify the holographic background. The requirement is that the field configuration is invariant under the Euclidean group and the dilatation. In the Lorentzian signature, it has been studied in \cite{Nakayama:2009qu} that such backgrounds are impossible in the pure geometric setup, and a possible violation from the flux is also forbidden by demanding the null energy condition \cite{Nakayama:2009fe}. Futhermore, the embedding of such field configurations in the warped compactification of the higher dimensional supergravity is also forbidden \cite{Nakayama:2009fe}. This is consistent with the (conjectured higher dimensional generalization of) the Zamolodchikov-Polchinski theorem, which states that scale invariant field theories are automatically conformal invariant under the assumptions of Poincar\'e invariance, unitarity, and the discreteness of the spectrum.

However, what we are interested in here is intrinsically Euclidean, and there is no reason at all why we have to impose the unitarity condition (or reflection positivity after Wick rotation). Indeed, even the simplest free example of $W[S(x)]$ proposed by Bialek and Zee is non-unitary and hence avoids the Zamolodchikov-Polchinski theorem. It is, therefore, natural to expect that Euclidean gravity systems might admit such a scale invariant but non-conformal field configuration.

In the Euclidean gravity there is no notion of energy condition, so it is possible to obtain the field configuration that is scale invariant but not conformally invariant.\footnote{However, it is easy to see that the pure geometry without compactification cannot support scale invariant but non-conformal background.} We will show one particular example based on the vector condensation. We claim that the Lorentzian continuation is necessarily non-trivial, and furthermore in  \cite{Nakayama:2009qu}, we have argued that such a continuation would result in an inconsistent background as a quantum gravity system with a holographic interpretation. It would simply contradict with the above-mentioned field theory theorem at least in $(1+2)$ dimensional bulk.
We are not going to consider the Lorentzian continuation in the following, but we would come back to the issue at the end of the paper.

A simple toy model of the scale invariant but non-conformal bulk field configuration is obtained in a Euclidean version of the vector condensation model in three-dimension (see \cite{Nakayama:2009qu} for the corresponding Lorentzian solution). The Euclidean action is given by
\begin{align}
S = \int d^{3}x \sqrt{g} \left( \frac{1}{2} \mathcal{R} - \Lambda + \frac{1}{4}\mathcal{F}_{\mu\nu} \mathcal{F}^{\mu\nu} + \sum_{n=1} \frac{g_n}{2n}(\mathcal{A}^\mu \mathcal{A}_\mu)^n \right)\ , \label{aE}
\end{align}
where $\mu = 1,2, z$, and the field strength is defined as $\mathcal{F}_{\mu\nu} = \partial_\mu \mathcal{A}_\nu - \partial_\nu \mathcal{A}_\mu$. We take a particular solution of the equation of motion
\begin{align}
ds^2 &= \frac{dx_1^2 + dx_2^2 + dz^2}{dz^2} \cr
\mathcal{A_\mu} dx^\mu &= \frac{a dz}{z} \ ,
\end{align}
where $a$ is determined by the coupling constants $g_n$ in \eqref{aE}.

The metric and the vector field are invariant under the dilatation
\begin{align}
x_{i} \to \lambda x_{i} \ , \ \  z \to \lambda z \ 
\end{align}
with $i=1,2$ as well as the Euclidean group of the $(1,2)$ plane. However, for $a\neq 0$, the vector field is not invariant under the special conformal transformation:
\begin{align}
\delta x^i = 2(\epsilon^i w_i)w^i - (z^2 +w^iw_i)\epsilon^i \ , \ \ \delta z = 2(\epsilon^i w_i) z \ . 
\end{align}
Therefore, the vector condensation model is an example of the bulk field configuration that is scale invariant but not conformally invariant. We note that the field configuration is locally stable in the Euclidean signature.

As we have discussed, the human perception is invariant under the dilatation but not under the special conformal transformation, so our vector condensation model may be used as a candidate for the holographic dual of the perception. On the other hand, in the Euclidean gravity, unlike the Lorentzian quantum gravity where they are scarce, we expect many more complicated models that admit scale invariant but non-conformal field configurations. In section 4, we use this toy model to present some explicit computations of the Bayesian probability distribution for perception from the holography. In the rest of the section, however, we develop the model-independent formalism of computing the Bayesian probability distribution by using the holographic technique.

\subsection{GKPW partition function $=$ Bayesian distribution function}
As we reviewed in section 2, the central problem in the Bialek-Zee model is to compute the Bayesian probability distribution encoded in the Euclidean effective action
\begin{align}
e^{-S_{\text{eff}}[\phi(x);\phi_0(x)]} = \int \mathcal{D}S(x) \exp\left(-W[S(x)] - \frac{1}{2C} \int d^2x \left(\phi(x)-\phi_0(y(x))\right)^2 \right) \ , \label{effaa}
\end{align}
where $y_i(x) = x_i+ S_i(x)$. In this subsection, we study the perturbative expansion of \eqref{effaa} with respect to the deviation $\Delta \phi = \phi - \phi_0$, and see how the Bayesian probability function is related to the GKPW partition function \cite{Gubser:1998bc}\cite{Witten:1998qj} of the dual gravity system.

We expand the image function with respect to $S_i(x)$ as
\begin{align}
\phi_0(y(x)) = \phi_0(x) + S_i(x) \partial_i\phi_0(x) + \cdots \ . \label{expa}
\end{align}
One may regard it as a derivative expansion to obtain a local functional $S_{\text{eff}}[\phi(x);\phi_0(x)]$. At the first order in $\Delta \phi$, we see that the Baysian probability distribution function can be expressed as
\begin{align}
e^{-S_{\text{eff}}[\phi(x);\phi_0(x)]} \sim \int \mathcal{D}S(x) \exp\left(-W[S(x)] - \frac{1}{2C} \int d^2x \left( \Delta \phi_0(x)^2 - 2\Delta\phi(x)\partial_i \phi_0(x)S_i(x) \right) \right) \ . \label{effab}
\end{align}

At this point, we can essentially regard the Bayesian probability distribution function as a generating functional for the correlation functions with the source term $\int d^2x \mathcal{J}_i(x) S_i(x)$ with $\mathcal{J}_i(x) = \frac{1}{C}\Delta \phi(x) \partial_i \phi_0(x)$. Moreover, in the GKPW prescription of the holographic computation of the correlation functions, the generating functional is identified with the partition function of the dual gravitational system with the specified boundary condition:
\begin{align}
Z[\mathcal{J}_i] &= \int \mathcal{D} S_i(x) \exp\left(-W[S(x)] +\int d^2x \mathcal{J}_i(x) S_i(x) \right) \cr
       &= Z_{\text{GKPW}}[\mathcal{J}_i] \cr
       &= \int \mathcal{D} g_{\mu\nu} \mathcal{D}\mathcal{A}_\mu|_{\mathcal{A}_i(z=0) = \mathcal{J}_i} \exp\left(-S_{\text{grav}}[g_{\mu\nu}, \mathcal{A}_\mu]\right) \ . \label{key}
\end{align}

The three-dimensional gravitational action $S_{\text{grav}}[g_{\mu\nu}, \mathcal{A}_\mu]$ does depend on $W[S(x)]$, and in general, we except that when $W[S(x)]$ is more complicated and difficult to integrate, the gravitational description becomes more tractable. As we have discussed, the scale invariant but non-conformal nature of $W[S(x)]$ demands that $S_{\text{grav}}[g_{\mu\nu}, \mathcal{A}_\mu]$ should admit a scale invariant but non-conformal bulk field configuration as studied in the previous subsection. To obtain the original Bayesian probability distribution, we simply replace $\mathcal{J}_i$ with $\Delta \phi(x) \partial_i \phi_0(x)$ and multiply it by $\exp\left(-\frac{1}{2C} \int d^2x \Delta\phi_0(x)^2 \right)$ in this approximation.

At the next order of approximation, we find the mass-like two-particle source term in the action:
\begin{align}
\frac{1}{2C} \int d^2x S_i(x) S_j(x) \partial_i\phi_0(x) \partial_j \phi_0(x) \ . \label{tso}
\end{align}
A perturbative treatment of this contribution from the holographic approach goes as follows. We introduce the dual source term for the symmetric tensor operator $S_i(x) S_j(x)$ by $\mathcal{J}_{ij}$. In the large $N$ limit of the gauge theory, the source $\mathcal{J}_{ij}$ is not independent of $\mathcal{J}_i$ but rather it is given by a two-particle source from $\mathcal{J}_i$. In the strongly coupled case, the non-perturbative effects may generate a tensor bound state for $S_iS_j(x)$ and $\mathcal{J}_{ij}$ may be treated effectively as and independent source from $\mathcal{J}_i$.

A generalization of the key relation \eqref{key} gives
\begin{align}
Z[\mathcal{J}_i,\mathcal{J}_{ij}] &= \int \mathcal{D} S_i(x) \exp\left(-W[S(x)] +\int d^2x \mathcal{J}_i(x) S_i(x) + \mathcal{J}_{ij} S_iS_j(x) \right) \cr
       &= Z_{\text{GKPW}}[\mathcal{J}_i,\mathcal{J}_{ij}] \cr
       &= \int \mathcal{D} g_{\mu\nu} \mathcal{D}\mathcal{A}_\mu|_{\mathcal{A}_i(z=0) = \mathcal{J}_i}\mathcal{D}\mathcal{T}_{\mu\nu}|_{\mathcal{T}_{ij}(z=0) = \mathcal{J}_{ij}} e^{-S_{\text{grav}}[g_{\mu\nu}, \mathcal{A}_\mu,\mathcal{T}_{ij}]} \ .
\end{align}
Here $\mathcal{T}_{\mu\nu}$ is a symmetric tensor field corresponding to a bound state for a pair of ``particles" made out of $\mathcal{A}_\mu$.
In this expression, we have treated $\mathcal{T}_{\mu\nu}$ as an independent field with $\mathcal{A}_\mu$. In the weakly coupled case, the constraint must be imposed appropriately in the definition of the bulk path integral.
In this way, the perturbative expansion of the GKPW partition function of the dual gravity  $Z_{\text{GKPW}}[\mathcal{J}_i,\mathcal{J}_{ij}]$ reproduces the Bayesian probability distribution by substituting $\mathcal{J}_i =  \frac{1}{C}\Delta \phi(x) \partial_i \phi_0(x)$ and $\mathcal{J}_{ij} = \frac{1}{2C} \partial_i\phi_0(x) \partial_j \phi_0(x)$.

An alternative approach to treat the two-particle source \eqref{tso} was proposed in \cite{Bialek:1986it}, where they replaced the derivative $\partial_\mu \phi_0(x) \partial_\nu \phi_0(x)$ by a constant $\delta_{\mu\nu} \phi_0^2/l^2$, where $\phi_0$ and $l^{-1}$ are typical values of $\phi(x)$ and its logarithmic derivative. In the free field theory model, the effect is to make the propagator for $S_i$ massive:
\begin{align}
\langle S_i(k) S_j(p) \rangle \sim \delta(k+p)\frac{1}{(k^2+\phi_0^2l^{-2}) \delta_{ij} + ck_ik_j} \ . \label{simple}
\end{align}
Accordingly, we expect that the leading long-range behavior of the Bayesian probability distribution is expotentially damping $\sim e^{-|x-y|/\xi}$.

In the dual gravity approach, the above approximation is equivalent to the deformation that makes the field theory massive. In the AdS/CFT correspondence, such massive deformations of the $\mathcal{N}=4$ theory down to $\mathcal{N}=2^*$ or $\mathcal{N}=1^*$ theory have been studied in the literatures. The correlation functions are not as simple as \eqref{simple} because the infrared physics is strongly coupled. We expect that the massive deformations of the scale invariant but non-conformal geometry is possible, and the Bayesian probability distribution can be computed as a partition function for such a deformed geometry by a suitable generalization of the GKPW prescription in massive backgrounds:

Generalizations to higher order in the derivative corrections are obvious. We can introduce the additional source term $\mathcal{J}_{ijk \cdots}$ corresponding to the higher derivative expansions in \eqref{expa}. The GKPW prescription dictates that we have the corresponding fields $\mathcal{T}_{\mu\nu\rho \cdots}$ in the gravitational system. We now compute the GKPW partition function $Z[\mathcal{J}_{ijk \cdots}]$ and identify it as the Bayesian probability distribution. The identification of $\mathcal{J}_{ijk \cdots}$ is also affected by the higher order expansion. For instance, the first order identification $\mathcal{J}_{ij} = \frac{1}{2C} \partial_i\phi_0(x) \partial_j \phi_0(x)$ obtains a correction $\delta\mathcal{J}_{ij} = \frac{1}{2C} \Delta\phi \partial_i\partial_j \phi_0$ at the next order.

Finally, let us briefly mention the prior distribution functional $e^{-S[\phi_0(x)]}$, and its integration over a measure $\mathcal{D}\phi_0(x)$ to obtain a posteriori probability $P[\phi_0(x)|\phi(x)]$. Again, one can regard the functional integral
\begin{align}
\int \mathcal{D} \phi_0(x) \exp\left(-S_{\text{eff}}[\phi(x);\phi_0(x)]-S[\phi_0(x)]\right)
\end{align}
as a path integral over the Euclidean field $\phi_0(x)$. It is not always easy to guess the prior distribution, and Bialek and Zee implicitly chose the flat prior distribution $S[\phi_0] = \text{const}$. Another natural choice would be $S[\phi_0(x)] = \alpha^{-1} \int d^2x \partial_i \phi_0(x) \partial_i \phi_0(x)$. The choice is based on our prejudice that the natural image is likely to be smooth so that the true image is more likely reproduced after averaging the gradient of the noise. This is similar to the spirit to use the Ising model for the Bayesian image restoration. The Gaussian functional integral over $\phi_0(x)$ is again a non-trivial task to perform, but the holographic approach might be useful here as well.

\section{Sample computation}
Now, we would like to show sample computations of the Bayesian probability distribution by using the holographic technique. In this section, we use the background studied in section 3.1 to implement the strategy discussed in section 3.2. 

We first have to identify the gravity dual of the vector field $S_i$. One candidate is the fluctuation of the vector field $\mathcal{A} = a\frac{dz}{z} + A_\mu dx^\mu$ around the background $\mathcal{A} = a\frac{dz}{z}$ appearing in the vector condensation model. Alternatively, one can also introduce an independent vector field $A_\mu$ that couples with the background $\mathcal{A}$.

The effective dynamics of a vector field in the vector condensate is governed by the Euclidean action
\begin{align}
S = \int d^{d+1} x \sqrt{g} \left(\frac{1}{4}F_{\mu\nu}F^{\mu\nu} + \frac{1}{2} m^2 A_i A^i + m_0^2 A_zA^z \right) \ , \label{actionf}
\end{align}
where $i=1,2,\cdots d$, and we assume that the metric is given by the AdS space:\footnote{We will keep the dimensionality $d$ of the space  arbitrary in the following discussion to make the computation slightly more general. In our application, we can simply set $d=2$.}
\begin{align}
ds^2 = \frac{dz^2 + dx_i^2}{z^2} \ .
\end{align}
Possible self-interaction terms of the vector field are omitted in order to focus on the two-point function. The difference between the ``mass" for $A_z$ and that for $A_i$ is crucial in the following discussion. For instance, it is generated by the term $(\mathcal{A}^\mu \mathcal{A}_\mu)^2$ in \eqref{aE} that includes the term $\mathcal{A}^z A_z \mathcal{A}^z A_z$.
It is this difference that breaks the conformal invariance or AdS isometry perceived by the vector field $A_\mu$. Our following computation is a generalization of that in \cite{Mueck:1998iz} for $m_0 \neq m$,

The equations of motion from \eqref{actionf} 
\begin{align}
D^\mu F_{\mu i} -m^2 A_i = 0 \ , \ \ D^\mu F_{\mu z} - m^2_0 A_z = 0 
\end{align}
imply the subsidiary condition
\begin{align}
m_0^2 \partial_z (z^{-d+1} A_z) + m^2 z^{-d+1} (\partial_i A_i) = 0 \ . \label{aux}
\end{align}
By using the subsidiary condition, we rewrite the equations of motion as
\begin{align}
\frac{m_0^2}{m^2}z^2 \partial^2_z A_z + z^2\partial_i^2 A_z - \frac{m_0^2}{m^2} (1-d)z\partial_z A_z - \left(m_0^2 - \frac{m_0^2}{m^2}(d-1)\right)A_z = 0 \ \label{first}
\end{align}
and
\begin{align}
&z^2 \left(\partial_z^2 + \partial_j^2\right)A_i + (3-d) z \partial_z A_i - m^2 A_i \cr 
=& z^2\left(1-\frac{m_0^2}{m^2}\right) \partial_z \partial_i A_z + z \left(\frac{m^2_0}{m^2}(1-d) - (d-3) \right) \partial_i A_z  \ . \label{second}
\end{align}

The first equation \eqref{first} can be solved by assigning the boundary condition $A_z \to 0$ as $z \to \infty$ with the usage of the modified Bessel function:
\begin{align}
A_z = \int \frac{d^d k}{(2\pi)^d} e^{i kx} a_0(k) z^{\frac{d}{2}} K_{\tilde{\alpha}}\left(k\frac{m_0}{m} z \right) \ ,
\end{align} 
where $\tilde{\alpha} = \sqrt{\frac{(d-2)^2}{4} + m^2}$.
The homogeneous part of the solution of the second equation \eqref{second} is given again by the modified Bessel function with a different argument:
\begin{align}
A_i = \int \frac{d^d k}{(2\pi)^d} e^{i kx} a_i(k) z^{\frac{d}{2}-1} K_{\tilde{\alpha}}\left(kz \right) + (\text{inhomogeneous}) \ .
\end{align}
When $m_0=m$, the inhomogeneous part also has an explicit solution:
\begin{align}
 \int \frac{d^d k}{(2\pi)^d} e^{i kx} ia_0(k)\frac{k_i}{k} z^{\frac{d}{2}} K_{\tilde{\alpha}+1}\left(kz \right)\ , \label{a0}
\end{align}
 but in a general case it seems difficult to obtain a closed form of the solution expressed in terms of known analytic functions, so we restrict ourselves to the power series solution.\footnote{As we will discuss at the end of this section, we can write down the explicit form of the inhomogeneous solution as an integral of Bessel functions, but it is less useful for the purpose of  determining the two-point function.}

As in \cite{Mueck:1998iz}, we introduce the fields with tangent space indices by
\begin{align}
\tilde{A}_a = e^\mu_a A_\mu = z \delta_{\mu a} A_\mu \ 
\end{align}
for $a = 1,2, \cdots, d,z$. Without confusion, we use the same indices $i = 1,2,\cdots,d$ in $\tilde{A}_i$.
Now we assume that $\tilde{A}_i$ can be expanded as
\begin{align}
\tilde{A}_i =& \int \frac{d^d k}{(2\pi)^d} e^{i kx} \left( a_j(k)\left(\delta_{ij} - \frac{k_ik_j}{k^2} \right) z^{\frac{d}{2}} K_{\tilde{\alpha}}(kz) \right. \cr
 &+ \left. i k_i \left(\sum_{n=0}^{\infty} c_{n}(k) z^{\frac{d}{2}-\tilde{\alpha} + 2n} + \tilde{c}_{n}(k) z^{\frac{d}{2} + \tilde{\alpha} + 2n} \right) \right) \ . \label{texp}
\end{align}
We naturally demand that $a_i(k)$ satisfies the transversality condition $k_i a_i(k)=0$. Once we fix $c_0(k)$, $\tilde{c}_0(k)$ and $a_0(k)$, the equations of motion \eqref{second} as well as the subsidiary condition \eqref{aux} determine $c_n(k)$ and $\tilde{c}_n(k)$ for $n>0$. In particular, note that $a_0(k)$ only affects the terms with $n>0$. 

More precisely, the equation of motion at the second order demands that
\begin{align}
c_1(k) =& \frac{k^2 c_0(k)}{4-4\tilde{\alpha}} \cr 
+ & \frac{a_0(k)}{4-4\tilde{\alpha}}\frac{\Gamma(\tilde{\alpha})}{2} \left(\frac{k}{2}\frac{m_0}{m}\right)^{-\tilde{\alpha}}\left(\left(1-\frac{m_0^2}{m^2}\right)\left(\frac{d}{2}-\tilde{\alpha}\right) + \frac{m_0^2}{m^2} (1-d) - (d-3) \right) \ 
\end{align}
and similarly for $\tilde{c}_1(k)$. On the other hand, the subsidiary condition determines $c_0(k)$ and $\tilde{c}_0(k)$ with respect to $a_0(k)$:
\begin{align}
c_0(k) &= \left(-\frac{d}{2}+1-\tilde{\alpha}\right) \frac{m_0^2}{k^2m^2} \frac{\Gamma(\tilde{\alpha})}{2} \left(\frac{k m_0}{m}\right)^{-\tilde{\alpha}} a_0(k) \cr
\tilde{c}_0(k) &= \left(-\frac{d}{2}+1+\tilde{\alpha}\right) \frac{m_0^2}{k^2m^2} \frac{\Gamma(-\tilde{\alpha})}{2} \left(\frac{k m_0}{m}\right)^{\tilde{\alpha}} a_0(k) 
\end{align}
The wavefunction $a_i(k)$ can be determined by demanding the Dirichlet boundary condition at $z = \epsilon$ as
\begin{align}
a_i(k) = \frac{\tilde{A}_i(k)}{K_{\tilde{\alpha}}(k\epsilon) \epsilon^{\frac{d}{2}}} \ , \label{dira}
\end{align}
where $\tilde{A}_i(k)$ is the Fourier transform of $\tilde{A}_i(x,z=\epsilon)$. 
To determine $c_0(k)$ and $\tilde{c}_0(k)$, we first write $\tilde{c_0}(k) = \lambda(k) c_0(k)$ to abbreviate the notation, where 
\begin{align}
\lambda(k) = \left(\frac{km_0}{2m}\right)^{2\tilde{\alpha}} \frac{\Gamma(-\tilde{\alpha})}{\Gamma(\tilde{\alpha})}\left(\frac{\frac{d}{2}-\tilde{\alpha}-1}{\frac{d}{2}+\tilde{\alpha} - 1} \right) \ .
\end{align}
  In the following, we will only need the first non-trivial expansion of $c_0(k)$ so that we expand it as
\begin{align}
c_0(k) = x_0(k) \epsilon^{-\frac{d}{2}-\tilde{\alpha}} + y_0(k) \epsilon^{-\frac{d}{2}+\tilde{\alpha}} + O(\epsilon^{-\frac{d}{2}-\tilde{\alpha} +2}) \ .
\end{align}
In the same approximation, the Dirichlet boundary condition at $z=\epsilon$ gives 
\begin{align}
x_0(k) = \frac{k_i\tilde{A}_i(k)}{ik^2} \ , \ \ y_0(k) = \frac{-k_i \tilde{A}_i(k)}{ik^2} \lambda(k) \ . \label{samea}
\end{align}

We use the standard AdS/CFT recipe to compute the two-point functions of the vector operator $\mathcal{J}_i$ that couples with the vector fields $A_i$. After integration by parts and using the equations of motion, the Euclidean action \eqref{actionf} becomes 
\begin{align}
S = -\frac{1}{2} \int d^d x \epsilon^{-d} \tilde{A}_i(\epsilon) \left(-\tilde{A}_i(\epsilon) + \epsilon \tilde{F}_{zi}(\epsilon) \right) \ , 
\end{align}
where $\tilde{F}_{zi} = \partial_z \tilde{A}_i - \partial_i \tilde{A}_z$. We actually need only the contribution from $\partial_z \tilde{A}_i(\epsilon)$ in the following because $\partial_i \tilde{A}_z$ is $O(\epsilon^2)$ smaller and can be neglected in the computation of the boundary two-point function.

In the approximation needed in our computation, the combination of \eqref{texp}, \eqref{dira} and \eqref{samea} yield
\begin{align}
\tilde{F}_{zi} = \int \frac{d^d k}{(2\pi)^d} e^{i kx} &\left[ \left(\frac{d}{2}-\tilde{\alpha}\right)\frac{1}{\epsilon} \tilde{A}_i(k)  \right. +2\frac{\Gamma(-\tilde{\alpha})}{\Gamma(\tilde{\alpha})}\tilde{\alpha} \left(\frac{k}{2}\right)^{2\tilde{\alpha}} \epsilon^{2\tilde{\alpha}-1} \left(\delta_{ij} -\frac{k_ik_j}{k^2} \right) \tilde{A}_j(k) \cr 
&+ \left. 2\tilde{\alpha} \lambda (k)\epsilon^{2\tilde{\alpha}-1} \frac{k_ik_j}{k^2} \tilde{A}_j(k) \right] \ .  
\end{align}
Accordingly, by discarding the contact terms,\footnote{The importance of the contact terms in these expressions in the context of threshold corrections in the renormalization group flow was discussed in \cite{Ho:2009zv}.} the two-point functions in the momentum space can be computed as
\begin{align}
\left\langle \mathcal{J}_i(k) \mathcal{J}_j(p)\right\rangle = C\delta^{(d)}(k-p)k^{2\tilde{\alpha}}\left( \delta_{ij} -\frac{k_ik_j}{k^2} + \lambda(k) \frac{\Gamma(\tilde{\alpha})}{\Gamma(-\tilde{\alpha})} \left(\frac{k}{2}\right)^{-2\tilde{\alpha}} \frac{k_ik_j}{k^2} \right)  \ 
\end{align}
with a constant factor $C$. Substituting the explicit value of $\lambda(k)$, we finally obtain
\begin{align}
\left\langle \mathcal{J}_i(k) \mathcal{J}_j(p)\right\rangle &= C\delta^{(d)}(k-p)
k^{2\tilde{\alpha}}\left( \delta_{ij} - \left(\frac{m_0}{m}\right)^{2\tilde{\alpha}} \frac{2\tilde{\alpha}}{\frac{d}{2}+\tilde{\alpha}-1}\frac{k_ik_j}{k^2}\right) \ . \label{corns}
\end{align}
The expression agrees with the most general two-point functions of vector operators in scale invariant but not necessarily conformally invariant field theories. In particular, when $m_0 = m$, the two-point function is conformally invariant \cite{Mueck:1998iz}.

In unitary conformal field theories, a conserved current always has a dimension $d-1$ and the converse is true: the primary vector fields that have a dimension $d-1$ are always conserved. In the scale invariant field theory, neither is guaranteed. Indeed, the holographic correlation functions \eqref{corns} suggests that a conserved current with $\Delta \neq d-1$ is possible by choosing $\left(\frac{m_0}{m}\right)^{2\tilde{\alpha}} = \frac{\frac{d}{2}+\tilde{\alpha} -1}{2\tilde{\alpha}}$. 

In order to compute the tree-level higher point correlation functions, we need boundary-bulk propagators which will be integrated over the volume of the bulk geometry. As noticed in \cite{Mueck:1998iz}, in the higher point correlation functions, we do not have to keep the non-analytic part of the bulk solutions of $A_\mu(k)$. For instance, it is enough to keep the leading term in
\begin{align}
A_z^{\text{bulk}} =  \int \frac{d^d k}{(2\pi)^d} e^{i kx} a_0(k)|_{\epsilon \to 0} z^{\frac{d}{2}} K_{\tilde{\alpha}}\left(k\frac{m_0}{m} z \right) \ .
\end{align}
By directly taking $\epsilon \to 0 $ limit in \eqref{a0} and \eqref{texp}, while keeping only the leading terms, we obtain the coefficients 
\begin{align}
a_0(k) &= \left(\frac{km_0}{m}\right)^{\tilde{\alpha}} \frac{2}{\Gamma(\tilde{\alpha})} \frac{k^2 m^2}{m_0^2} \frac{1}{-\frac{d}{2}+1-\tilde{\alpha}} \frac{k_i\tilde{A}_i}{ik^2 \epsilon^{\frac{d}{2}+\tilde{\alpha}}}  \cr
a_i(k) &= \frac{2}{\Gamma(\tilde{\alpha})} \frac{\tilde{A}_i}{k^{\tilde{\alpha}} \epsilon^{\frac{d}{2}+\tilde{\alpha}}}  \cr 
c_0(k) &= \frac{k_i\tilde{A}_i}{ik^2 \epsilon^{\frac{d}{2}+\tilde{\alpha}}} \cr
\tilde{c_0}(k) &= \lambda(k) c_0(k)
\end{align}
and so on for higher $c_n(k)$. It is tedious to find all the coefficients $c_n(k)$ in this approach. In the numerical computation of the bulk-boundary propagator, it is more direct to use the Green function for the differntial operator appearing the left hand side of \eqref{second}, which is essentially the Green function of the Bessel equation, to obtain the  integral representation for the inhomogeneous part of $\tilde{A}_i$ (see appendix B for the Green function and the solution of the inhomogeneous Bessel equation).

\section{Discussion}

In this paper, we have proposed a holographic model for invariant perception. It is based on the strongly coupled two-dimensional Euclidean field theory that is scale invariant but not conformal invariant. We have constructed the corresponding gravity configuration and studied the correlation functions. The GKPW partition function computed from the geometry is identified with the Bayesian probability distribution of the invariant perception through the generalization of the Bialek-Zee model.

We have studied only the classical holographic computation, and it remains open whether the model is consistent at the quantum level. Most probably the Euclidean gravity model (even in 1+2 dimension) is perturbatively ultraviolet divergent and non-renormalizable. Furthermore, the non-perturbative stability is always an important and delicate issue in the path integral appraoch to the Euclidean gravity. 

There are several possibilities to go beyond the classical computation. Here, we would like to  suggest some approaches that are available only to us and not applicable in the conventional AdS/CFT correspondence.
The one approach is to introduce the scalar or topological supersymmetry that would cancel the ultraviolet divergence. We do not necesserily have to introduce the ordinary spinor supersymmetry. A Euclidean field theory allows a scalar or vector supersymmetry. This is because the unitarity is from the begining abondoned in our Euclidean field theory setup, and the spin-statistics is irrelevant. Indeed, the scalar supersymmetry is ubiquitus in the random system or diffusion system, and the connection between the Baysian statistics and the random system is also of importance.

Another approach is to consider Euclidean string theories as an ultraviolet completion of the gravitational theory we have proposed. The Euclidean string theories possess much fewer physical states than the Lorentzian string theory, and the ``dynamics" is more restrained.\footnote{Essentially, the Virasoro constraint demands that only ``vacua" are physical states.} Again there is no reason to impose  spin-statistics or unitarity in the spectrum, so much more interesting structures could emerge. Solutions to such Euclidean string theories would provide holographic plethora of strongly coupled scale invariant Euclidean field theoires possibly without reflection positivity. We do not see any excuse not to study Euclidean stirng theories simply because they do not describe our entire universe.

\section*{Acknowledgements}
The author would like thank Heng-Yu Chen and Gary Shiu for stimulating conversations on the possiblity of AdS/Biology. 
The work was supported in part by the National Science Foundation under Grant No.\ PHY05-55662 and the UC Berkeley Center for Theoretical Physics. 

\appendix

\section{Scale invariance vs conformal invariance}
A local field theory is scale invariant when the trace of the symmetric energy momentum tensor is a total divergence:
\begin{align}
T^{i}_{ \ i} = -\partial^i J_i \ .
\end{align}
The current $J_i$ is known as virial current. One can construct the conserved dilatation current $D^i$ by
\begin{align}
D^i = x_jT^{ij} + J^i \ .
\end{align}
In the Lorentzian field theory, one can construct the dilatation charge out of $D^0$.

If the virial current is itself is a total divergence:
\begin{align}
T^{i}_{\ i} &= \partial_i \partial_j L^{ij} \ \ (d\ge 3) \cr
& = \partial^i \partial_i L \ \ \ (d=2) \ ,
\end{align}
one can improve the energy-momentum tensor so that it is traceless (see e.g. \cite{Polchinski:1987dy} for details)
\begin{align}
\Theta^{i}_{\ i } = 0 \ .
\end{align}

By using this improved traceless energy-momentum tensor, one can construct the conserved current:
\begin{align}
j^i_v = v_j \Theta^{ji} \ ,
\end{align}
where the vector $v_i$ satisfies
\begin{align}
\partial_i v_j + \partial_j v_i = \frac{2}{d} \delta_{ij} \partial_k v^k \ .
\end{align}
In the Lorentzian field theory, the charge associated with $j_v^i$ generates all the conformal transformation. Thus, the distinction between the scale invariance and the conformal invariance is reduced to the problem whether the virial current is a total derivative or not.

\section{Modified Bessel function}
We repeatedly use the modified Bessel functions in the computation of correlation functions in AdS/CFT correspondence (and its scale invariant cousins), so we will summarize the basic properties here. The modified Bessel function $K_\alpha(x)$ is a solution of the modified Bessel equation:
\begin{align}
x^2 \frac{d^2}{dx^2}K_\alpha(x) + x \frac{d}{dx}K_\alpha(x) -(x^2+\alpha^2)K_\alpha(x) = 0 \ , \label{modb}
\end{align}
and it decays exponentially $K_\alpha(x) \sim \sqrt{\frac{\pi}{2x}}e^{-x}$ as $x \to \infty$.

It has the integral form
\begin{align}
K_{\alpha}(x) = \frac{1}{2}e^{-\frac{1}{2}\alpha \pi i} \int_{-\infty}^{\infty} dt e^{-ix\sinh t - \alpha t }
\end{align}
and it can be expanded as 
\begin{align}
K_\alpha(x) =& \frac{1}{2}\Gamma(\alpha)\left(\frac{x}{2}\right)^{-\alpha}\left(1+\frac{x^2}{4(1-\nu)} + \frac{x^4}{32(1-\nu)(2-\nu)} + \cdots\right) \cr
&+ \frac{1}{2}\Gamma(-\alpha)\left(\frac{x}{2}\right)^{\alpha}\left(1+\frac{z^2}{4(1+\nu)} + \frac{x^4}{32(1+\nu)(2+\nu)} + \cdots\right) \ .
\end{align}
A useful recursion relation is
\begin{align}
\frac{\partial}{\partial z}K_\alpha(kz) = -kK_{\alpha-1}(kz) - \frac{\alpha}{z}K_\alpha(kz) \ .
\end{align}

The inhomogeneous modified Bessel equation
\begin{align}
x^2 \frac{d^2}{dx^2}G(x) + x \frac{d}{dx}G(x) -(x^2+\alpha^2)G(x) = f(x) \ 
\end{align}
can be solved by
\begin{align}
G(x) = I_\alpha (x)\int^{x} dz z^{-1} K_{\alpha}(z)f(z) - K_\alpha (x) \int^x dz z^{-1} I_\alpha (z)f(z) \ ,
\end{align}
where $I_\alpha(x) = \sum_{m=0}^\infty \frac{1}{m!\Gamma(m+\alpha +1)}\left(\frac{x}{2}\right)^{2m+\alpha}$ is another solution of the modified Bessel equation \eqref{modb}. The formula follows from the identity for the Wronskian of the modified Bessel equation:
\begin{align}
J_\alpha(x)\partial_x K_\alpha(x) - \partial_x J_\alpha(x) K_\alpha(x) = -x^{-1} \ .
\end{align}

\end{document}